
\magnification=\magstep1
\hoffset=0.1truecm
\voffset=0.1truecm
\vsize=23.0truecm
\hsize=16.25truecm
\parskip=0.2truecm
\def\newpage{\vfill\eject}

\def\fh1{ \bigl[- {d h \over dx} \bigr] }
\def\h2{ \bigl[- {1 \over h} {d h \over dx} \bigr] }
\def\f1{\Bigl[ \int_0^1 f dx \Bigr]}

%
\centerline{\bf CALCULATION OF
PARTICLE PRODUCTION BY NAMBU GOLDSTONE BOSONS
WITH APPLICATION TO INFLATION REHEATING AND BARYOGENESIS}
\bigskip
\centerline{\bf Alexandre Dolgov$^{1,2}$ and Katherine Freese$^1$}
\bigskip
\centerline{\it 1) Physics Department, University of Michigan}
\centerline{\it Ann Arbor, MI 48109, USA}
\bigskip
\centerline{\it 2) Permanent Address: ITEP, Moscow}
\centerline{\it 117259 Russia}
\bigskip
\centerline{\it submitted to Physical Review {\rm D}: October 4, 1994}

\vskip 0.40truein
\goodbreak
\centerline{\bf Abstract}
\medskip
A semiclassical calculation of particle production
by a scalar field in a potential is performed.
We focus on the particular case of production of fermions
by a Nambu-Goldstone boson $\theta$.
We have derived a (non)local equation of motion for the $\theta$-field
with the backreaction of the produced particles taken into
account. The equation is solved in some special cases, namely for
purely Nambu-Goldstone bosons and for the tilted potential
$ U(\theta ) \propto m^2 \theta^2 $.
Enhanced production of bosons due to parametric resonance
is investigated; we argue that the resonance probably disappears
when the expansion of the universe is included.
Application of our work on particle production
to reheating and an idea for
baryogenesis in inflation are mentioned.

\vskip 0.3truein
PAC Numbers: 98.80.--k, 98.80.Cq, 14.80.Mz, 11.30.Fs

\newpage
\centerline{\bf I. Introduction}
\medskip

Nambu-Goldstone bosons (NGBs) are ubiquitous in particle physics:
they arise whenever a symmetry is spontaneously broken.
If there is additional explicit symmetry breaking, these particles
become pseudo Nambu-Goldstone bosons (PNGBs).
In this paper, we consider particle production by Nambu-Goldstone
bosons as they rotate about the bottom of the `Mexican
hat' potential, whether or not there is a tilt around the
bottom (i.e. with or without explicit symmetry breaking).

The Nambu-Goldstone bosons, hereafter called $\theta$, are
assumed to couple to fermions; thus as the $\theta$ field moves it
is capable of producing these fermions.
Here we perform a semiclassical calcuation.
The $\theta$
field is treated classically while the particles produced are
quantized. The backreaction of
quantum fermions on the evolution of the $\theta$ field
is calculated.  Our motivation is to lay out a general
way to perform such a calculation and to carry it out
for a specific coupling.  The primary results of our calculations
in flat spacetime are Eqs. (2.21) and (2.22).  We then
demonstrate examples and generalize to curved spacetime
with massive fermions.

Particle production by Nambu Goldstone fields may have
several applications.  One is
the QCD axion.  Another is particle production by
the inflation field [1] in Natural Inflation [2].
Particle production is, of course, important for estimates
of reheating in inflation.
Many models of inflation involve `slowly rolling' fields
that evolve down a potential.  Subsequent to
the `slowly-rolling' epoch, there must be an epoch
of reheating, where vacuum energy is converted into
the production of radiation energy.
The equation of motion for the inflaton field is taken
to be
$$\ddot \theta + 3 H \dot \theta + \Gamma \dot \theta
= - {dU \over d\theta} \, . \eqno(1.1)$$
The term $\Gamma \dot \theta$ is assumed to describe
the reheating.  $\Gamma$ is taken to be the decay rate
of the inflaton field. This heuristic term
really describes much more complicated physics.  In fact
one should accurately calculate the production of
particles and its back reaction on the inflaton
field as it rolls down the potential.  Previous
work on this subject includes References [3,4].

In addition, there may be a mechanism
for baryogenesis during Natural Inflation.
If the equivalent
of the Peccei Quinn field can be made to carry baryon number,
one may be able to do baryogenesis as the inflaton is rolling
down its potential.  This has several nice features:
i) the same field would be responsible for both inflation
and baryogenesis and ii) the inflaton could reheat to very
low temperatures, perhaps as low as nucleosynthesis temperatures
of $\sim MeV$.  (In fact, for this mechanism to work one
would have to reheat to below the electroweak temperature
to avoid sphaleron destruction, if it is operative;
alternatively the inflaton
could generate nonvanishing $(B-L)$-asymmetry
which is preserved by sphalerons).
The approach is similar to proposals of Affleck
and Dine [5] and Cohen and Kaplan [6].
It is assumed that the inflaton field $\Phi$ is complex and
has a nonvanishing baryon
number.  The corresponding baryon current
generated by the classical rolling down of the inflaton field
is essentially equal to the angular momentum of the
two-dimensional mechanical motion in the plane $(Re\Phi, Im\Phi)$.
Thus, as the field
rolls in one direction, it preferentially creates baryons over
antibaryons, while the opposite is true as it rolls in the
opposite direction.  (We assume that the decays during reheating
are baryon number conserving).
Thus, no CP violation is required of the particle
physics; instead those regions of the universe in which
the inflaton by chance rolls down the potential
in one direction turn out to be baryon dominated, while those
that roll down the other direction turn out to be antibaryon
dominated.  Conveniently, each
of these regions is inflated
to be very large, so that it makes sense for our baryon dominated
region to be large enough to encompass our observable universe.
The requirements to create a specific particle physics
model for this proposal are restrictive and are discussed below.

In Section II, we discuss the sample Lagrangian we consider,
and calculate the particle production for this case.  First
we perform the calculation in the absence of expansion of the
universe, and for production of massless fermions.
In Section III, we apply the results to
two specific examples: i) a scalar $\theta$ rotating
in a potential without explicit symmetry breaking, and ii)
the same scalar but now
oscillating near the minimum of a quadratic potential
produced by explicit symmetry breaking; for example, such a potential
may give rise to inflation.  In Section IV, we include the effects
of expansion of the universe and fermion masses.
We also discuss the possibility of parametric resonance
[3,7,8] whereby large numbers of scalars might be produced during
reheating in inflation. We argue that there is probably no resonance
if one includes the expansion of the universe.
In Section V we discuss the possible application
to inflation, particularly to the model of Natural Inflation.
The baryogenesis model mentioned above is discussed in this section.
In Section VI we conclude.

\bigskip
\centerline{\bf II. Particle Production in a Simple Model}

\noindent
IIA) {\it Lagrangian:}  We first describe a simple model in
which we calculate particle production.  Consider the fundamental
action for a complex scalar field $\Phi$ and two fermions
$Q$ and $L$:
$${\it S} = \int d^4x \sqrt{-g} [g^{\mu\nu} \partial_\mu \Phi^*
\partial_\nu \Phi - V(\Phi^* \Phi) + i\bar Q \gamma^\mu \partial_\mu Q
+ i \bar L \gamma^\mu \partial_\mu L + (g \Phi \bar Q L + h.c.)]
\, . \eqno(2.1)$$
Note that $Q$ and $L$ can be any fermions, not necessarily
quarks and leptons of the standard model.  For example, they can
be heavy fermions; they may be given some of the same quantum numbers
as particles in the standard model if they couple to ordinary
quarks and leptons.  For this section of the paper, we will
take the intrinsic mass of the $Q$ and $L$ fields to be zero,
and will include mass effects in later sections.

This action is invariant under the appropriate $U(1)$ symmetry.
For example, in this paper, we will take the Lagrangian
to be invariant under
$$\Phi \rightarrow e^{i \alpha} \Phi, ~~~~Q \rightarrow e^{i \alpha} Q,
{}~~~~L \rightarrow L \, . \eqno(2.2a)$$
Eq. (2.2a) is the symmetry we will use for the rest of the paper.

We did, however, want to point out that a very similar analysis
would apply to the case of global chiral $U(1)$ symmetry
in a Lagrangian with Yukawa coupling $g \bar \psi_L \psi_R \Phi$.
Here subscripts $L$ and $R$ refer to left- and right-handed
projections of the fermion fields, $\psi_{R,L} = (1 \pm \gamma_5)
\psi/2$.
This Lagrangian is invariant under
$$\psi_L \rightarrow e^{i \alpha/2} \psi_L, ~~~~\psi_R \rightarrow
e^{-i \alpha/2}, ~~~~
\Phi \rightarrow e^{i \alpha} \Phi \, , \eqno(2.2b)$$
which is the Peccei-Quinn (PQ) symmetry [9] in axion models.
Then the current would be $J^\mu = \bar \psi \gamma^\mu
\gamma^5 \psi$.
Although we do not explicitly analyze this case, it would
be very similar to the one we do look at.

We assume the global symmetry is spontaneously broken at the
energy scale $f$ in the usual way, e.g. via a potential of the
form
$$ V(|\Phi|) = \lambda \Bigg[\Phi^* \Phi - f^2/2 \Bigg]^2 \, .
\eqno(2.3)$$
The resulting scalar field vacuum expectation value (VEV)
is $\langle \Phi \rangle = f e^{i\phi/f}/\sqrt{2}$.

Below the scale $f$, we can neglect the superheavy radial mode of
$\Phi$ ($m_{radial} = \lambda^{1/2} f$) since it is so
massive that it is frozen out.  The remaining light degree of freedom
is the angular variable $\phi$, the Goldstone boson of the
spontaneously broken $U(1)$ (one can think of this as the angle around
the bottom of the Mexican hat described by eqn. (2.3)).
For simplicity of notation, we introduce the dimensionless
angular field $\theta \equiv \phi/f$.
We thus study
the effective Lagrangian [10] for $\theta$:
$${\cal L}_{eff}
= {f^2\over 2} \partial_\mu \theta \partial^{\mu} \theta + i \bar Q
\gamma^{\mu} \partial_\mu Q + i \bar L
\gamma^{\mu} \partial_\mu L +
 (gf \bar Q L e^{i\theta} + h.c.) - U(\theta)
{}~. \eqno(2.5)
$$

The global symmetry is now realized
in the Goldstone mode: ${\cal L}_{eff}$ is
invariant under
$$ Q \rightarrow e^{i \alpha} Q, ~~~~L \rightarrow L,
{}~~~~\theta \rightarrow \theta + \alpha  \, .\eqno(2.6)$$
At this stage, $\theta$ is massless because we have not yet
explicitly broken the symmetry.

With a rotation of the form in Eq. (2.6) with $\alpha = - \theta$,
the Lagrangian can alternatively be written
$${\cal L}_{eff}
= {f^2\over 2} \partial_\mu \theta \partial^{\mu} \theta + i \bar Q
\gamma^{\mu} \partial_\mu Q + i \bar L
\gamma^{\mu} \partial_\mu L + (gf \bar Q L + h.c.) + \partial_\mu
\theta J^\mu - U(\theta) ~, \eqno(2.7)
$$
where the fermion current derives from the $U(1)$ symmetry;
here, $J^\mu = \bar Q \gamma_\mu Q$.

{\it Explicit symmetry breaking:}  Our subsequent analysis
of particle production applies whether or not the symmetry
is further broken explicitly.  Several options
exist for explicitly breaking the global symmetry and generating
a PNGB potential at a mass scale $\sim \Lambda$.  Models
include the the schizon models of [Ref. 12].  Another
possibility is the QCD axion [13]: dynamical chiral symmetry breaking
through strongly coupled gauge fields. When QCD becomes strong
at a scale $\Lambda_{QCD} \sim GeV$, instanton effects become
important.  Chiral dynamics induces a fermion condensate,
$\langle\bar \psi \psi \rangle \sim \Lambda^3$, and the potential for the
angular PNGB field becomes
$$U(\theta) = \Lambda^4 [1\pm {\rm cos}\theta] \, . \eqno(2.8)$$
Such a potential is used in the case of Natural Inflation,
although at higher mass scales, and will be discussed later.

{\it Equations of Motion:}  The equations of motion for the
$Q$ fields are
$$ i \gamma^\mu \partial_\mu Q + \partial_\mu \theta \gamma^\mu Q
 + gf L = 0 \,  \eqno(2.9a) $$ and
$$i \partial_\mu(\bar Q \gamma^\mu) - \partial_\mu \theta  \bar Q
\gamma^\mu - g f \bar L = 0 \, . \eqno(2.9b)$$
The combination of these two equations can be written
$$\partial_\mu J_Q^\mu = \partial_\mu(\bar Q \gamma^\mu Q)
= -igf(\bar Q L - \bar L Q) \, . \eqno(2.10)$$
The equation of motion for $\theta$ is
$$ {\cal D}^2 \theta + U'(\theta)/f^2 = {-1 \over f^2} \partial_\mu
J^\mu = {ig \over f}(\bar Q L - \bar L Q) \, . \eqno(2.11)$$

The equation of motion for the $L$ field, which we assume
does not transform under the symmetry, is
$$i \gamma^\mu \partial_\mu L = - gf Q \, . \eqno(2.12)$$
As above, if we add the equation of motion for the $\bar L$
field, we find
$$\partial_\mu(\bar L \gamma^\mu L) = -igf(\bar Q L - \bar L Q)
\, . \eqno(2.13)$$

{\it Particle Production:}  Here we calculate production of
$Q$ and $L$ particles by the angular $\theta$ field as it
rotates around the Mexican hat (which may or may not be tilted).
For the moment we will neglect expansion of the universe.

For convenience we will define
$$Q = Q_0 e^{i \theta} \eqno(2.14)$$
so that eqn. (2.9) becomes
$$\gamma^\mu \partial_\mu Q_0 =i gf e^{-i \theta} L \, . \eqno(2.15)$$
We will solve perturbatively the Heisenberg equations of motion
presented above. We will take the free field
$Q_{in}$ to satisfy $\gamma^\mu \partial_\mu Q_{in} = 0$,
and make a perturbation expansion $Q_0 = Q_{in} + g Q_1$.
Eq. (2.15) is solved by
$$Q_0(x) = Q_{in}(x) + i g f \int d^4y G_Q(x,y) L(y) e^{-i \theta(y)}
\, , \eqno(2.16a)$$
where to lowest order we will take $L = L_{in}$ inside the integral.
Here $G_Q$ is the
retarded Green's function for the $Q$ field
and satisfies $\partial_{\mu} \gamma^\mu G_Q(x,y) = \delta^4(x-y)$.
Similarly, the solution to Eq. (2.12) is
$$L(x) = L_{in}(x) + igf\int d^4y G_L(x,y) Q(y)  $$
$$ \approx L_{in}(x) + igf\int d^4y G_L(x,y) Q_{in}(y) e^{i \theta(y)}
\, . \eqno(2.16b)$$

We take the vacuum expectation value (in Heisenberg picture)
$$\langle{\cal D}^2 \theta(x) + U'(\theta(x))/f^2\rangle =$$
$${1\over f^2} \langle \partial_\mu J^\mu \rangle _{vac}
= {ig \over f} \langle  (\bar Q L - \bar L Q) \rangle_{vac} \, . \eqno(2.17)$$
We perform a semi-classical treatment, where the $\theta$
field is treated classically while the fermion fields are quantized
and via the equations of motion determine the evolution of $\theta$.
To first order in $g$, the right hand side of Eq. (2.17) is
$\langle\bar L_{in} Q_{in} \rangle = 0$.  Working to second order in $g$,
we substitute Eq. (2.16a) into Eq. (2.14), and plug that into
the right hand side of (2.17) to obtain
$$
\langle{\cal D}^2 \theta(x) + U'(\theta(x))/f^2\rangle =$$
$$
g^2\langle \int d^4y \bar L_{in}(x) G_Q(x,y) L_{in}(y) e^{-i \theta(y)}
e^{i \theta(x)} -
\bar Q_{in}(x) G_L(x,y) Q_{in}(y)
e^{i \theta(y)} e^{-i\theta(x)} + h.c. \rangle \, . \eqno(2.18)
$$
Here, $G_Q$ and $G_L$ are the Green's functions for the $Q$
and $L$ fields.  For the case of massless fermions, $G_L = G_Q$.

We now quantize the free fermion fields,
$$Q_{in}(x) = \sum_s \int{d^3k \over (2 \pi)^{3/2}} \sqrt{{m_Q \over E_k}}
[u_k^s b_k^s e^{-ik \cdot x} + v_k^s d_k^{s{\dagger} }
e^{ik \cdot x}] \, ,\eqno(2.19)$$
where $b_k^s$ and $d_k^{s{\dagger}}$ are annihilation and creation
operators at momentum $k$ and spin $s$ for particles and
antiparticles respectively.  The quantization for the free $L_{in}$
field is similar.  Now we evaluate
the right hand side of (2.17).  The details of the calculation
are discussed in Appendix A.  As shown there,
we find that
$$
\langle \bar L(x) G(x,y) L(y) \rangle =
4\int {d^3p \over (2 \pi)^7} \int {d^4l \over 2 E_p}
\Bigg[{i \over l^2 + m_Q^2}\Bigg]
 e^{[i(p+l) \cdot (y-x)]} l \cdot p \, , \eqno(2.20)$$
and a similar expression for the quarks.

After much algebra (presented in Appendix A), we find
$$
\langle {\cal D}^2 \theta + U'(\theta)/f^2 \rangle
= - {4g^2 \over \pi^2}
\int_0^\infty d \omega \omega^2 \int_{- \infty}^0 dt'
{\rm sin}(2 \omega t') {\rm sin}[\theta(t+t') - \theta(t)] . \eqno(2.21)
$$

As shown in Appendix A, after the $\omega$ integral is
done and with some algebra, this equation becomes
$$
\langle {\cal D}^2 \theta + U'(\theta)/f^2 \rangle =
- {g^2 \over 2\pi^2}
\lim_{w \rightarrow \infty}
\int_{-\infty}^0 dt' \Bigg[{{\rm cos}2 wt' - 1 \over
 t'}\Bigg] \Bigg[\ddot \theta(t+t') {\rm cos}\Delta \theta
- \dot \theta^2(t+t') {\rm sin}\Delta \theta\Bigg] \, . \eqno(2.22)
$$
Here $w$ tends to infinity and
we have defined $\Delta \theta \equiv \theta(t+t') -
\theta(t)$.

Eqs. (2.21) and (2.22) are the major results of this section.
To reiterate, these equations describe the evolution of the
$\theta$ field with production of massless fermions taken
into account in a semiclassical approximation.  So far the
expansion of the universe has not been included.

\bigskip
\centerline{\bf III. Examples}

In this Section we will apply Eq. (2.21) to two examples.
First we will consider the case of $U'(\theta)=0$.  This
is the case where there is no explicit symmetry breaking.
Thus the potential looks like an ordinary Mexican hat,
in which every point around the bottom is equivalent.
For example, there is no tilt (no cosine potential).
Another situation in which this case would be relevant
is the case when the rotation proceeds
so far up
in the Mexican hat that the details around the bottom are
irrelevant and can be ignored.

The second example we will consider is one that would
be relevant to inflation, namely oscillations around
the bottom of a tilt in the potential.
In this case $U'(\theta) \neq 0$
and there is explicit breaking of the symmetry.

{\it Case I: $U'(\theta) = 0$:}
In this first case, we consider a simple Mexican hat
with no explicit symmetry breaking.
The zeroth order solution to Eq. (2.21) would be obtained
by setting the right hand side, which is proportional to $g^2$,
to zero.  The zeroth order solution is $\dot \theta = const$.
In other words, the field is simply rotating around in the
Mexican hat with constant angular velocity.
We substitute this ansatz $\dot \theta = const$,
back into Eq. (2.21) to see what
the corrections would be.
The $t'$ integral becomes
$\int_{-\infty}^0 dt' {\rm sin}(2 \omega t') {\rm sin}(\dot \theta t')
= {\pi \over 2}[\delta(2 \omega + \dot \theta) - \delta(2 \omega
-\dot \theta)]$.  Thus Eq. (2.21) becomes
$$
\ddot \theta + { g^2 \over 2\pi} \dot \theta^2 sign(\dot \theta) =0
\,. \eqno(3.1)
$$

If the field $\Phi$ has been arranged to carry baryon number,
then the baryon number is shifted (via a baryon conserving
decay) to the fermions.  The baryon number of the fermions
satisfies $\partial_{\mu} J^{\mu} = f^2 \ddot \theta$
from the equations
of motion. Thus
$\dot n_B = f^2 \ddot \theta$,
where $n_B$ is the baryon number carried by the fermions.
The change in baryon number
carried by the fermions is thus determined by the change
in $\dot \theta$ via
$\Delta n_B = f^2 \Delta \dot \theta$,
similar to Ref. [5].
Here, all that has happened is that this mechanism transfers
the initial baryon number $B_i = f^2 \dot \theta_i$ into
the quarks. Subscripts $i$ refer to initial values.
Thus it is an initial value problem: to get
the right value today one would need exactly the right value
of $\theta_i$.  We are not proposing this as a likely explanation
for the baryon content of our universe.

The solution to Eq. (3.1) is
$ \dot \theta = 2\pi \dot \theta_i /( 2\pi + g^2 t \dot \theta_i) $
and $\theta =
2\pi g^{-2}
\ln( 2\pi + g^2 t \dot \theta_i) + \theta_i$.  Here $\theta_i$ is
the initial value of $\theta$.  Thus, we have checked that
in the short time or small $g$ limit, $\dot \theta \sim const$ was
reasonable.  We wish, however, to remind the reader that
the expansion of the universe has not yet been included here.

{\it Case II: Oscillations around the Minimum of a Potential for $\theta$:}
Here will consider the case where $U'(\theta) \neq 0$,
i.e. there is a tilt as you go around the bottom of the Mexican hat.
As a simple example we will consider $U(\theta) =
m^2 f^2\theta^2/2 $,
as would be appropriate near the bottom of  a cosine potential.

Then Eq. (2.22) becomes
$$
\ddot \theta + m^2 \theta =
- {g^2 \over 2\pi^2} \lim_{w \rightarrow \infty}
\int_{-\infty}^0 dt'\Bigg[ {{\rm cos}2 w t' - 1 \over
 t'}\Bigg] \Bigg[\ddot \theta(t+t') {\rm cos}\Delta \theta
- \dot \theta^2(t+t') {\rm sin}\Delta \theta\Bigg]\, ,
\eqno(3.2)$$
Again, the zeroth order solution is obtained by setting the
right hand side equal to zero.
Thus as our ansatz we take $\theta(t) = \theta_0(t) {\rm cos}m_Rt$
where $m_R$ is the renormalized mass to be defined below.
We will assume that $\theta_0(t)$ varies more slowly with
time than do the cosine oscillations.
We will consider the case of small oscillations around the
bottom of the potential.  Then we can take cos$\Delta \theta
\approx 1$
and neglect $\dot \theta^2$ in Eq. (3.2).
[Note that the $t'$ integral should really start
from a finite time at which small oscillations begin;
here we will approximate the lower limit of the
integral as $t'_{initial} = - \infty$.]
Then the $t'$ integral
becomes
$$
-{g^2 \over 2 \pi^2}
\lim_{w \rightarrow \infty } \int_{-\infty}^0 d t'
\Bigg[{{\rm cos}2 w t' - 1 \over  t'}\Bigg] [-m^2_R \theta_0(t)
{\rm cos}m_R(t+t')] =$$
$$-{g^2 \over 2 \pi^2}
\lim_{w \rightarrow \infty} \Bigg[-m^2_R \theta_0 {\rm sin}m_Rt
\int dt' {{\rm sin} m_R t' \over  t'}
+ m^2_R \theta_0 {\rm sin}m_R t \int dt' {{\rm sin}m_R t'  \over  t'}
{\rm cos}2 w t'$$
$$+ m^2_R \theta_0   {\rm cos}m_R t \int
{dt' \over t'} (1 - {\rm cos}2 w t') {\rm cos}m_R t' \Bigg]
\, .
\eqno(3.3)$$
The first term
behaves like a friction term, $- g^2 m_R \dot \theta /(4 \pi) $.
The second term is zero after one takes the limit $w \rightarrow
\infty$.  The third term is a mass renormalization term.
After evaluating the integral in this third term
as shown in Appendix B, we find that this term is
${g^2 \over 2\pi^2} m_R^2 \theta  \log(2 w / m_R)$.
We wish to add this term to the second term on the
left hand side of Eq. [3.2] so that the sum of these
terms gives $m_R^2 \theta$.  Therefore we define
$m_R^2 \big[ 1 + {g^2 \over 2 \pi^2} \log(2 w / m_R)\big]
= m^2$.  Then the original integro-differential equation
is effectively reduced to
$$\ddot \theta + m_R^2 \theta + \Gamma \dot \theta = 0 \, , \eqno(3.4)$$
where $\Gamma \equiv g^2 m_R /4\pi$.
The solution to this equation is
$$
\theta(t) = \theta_i e^{- \Gamma t /2} {\rm cos}(m_R t + \delta)\, .
\eqno(3.5)$$
where we introduced an arbitrary phase $\delta$ which is fixed by
initial conditions.

Equation (3.4) (or (1.1)) describes the damping of the external
field oscillations due to particle production. It was postulated
in many papers where the universe reheating was considered.
As we have shown it is indeed correct, but our approach
alerts us to several issues that should be considered
further with regard to the calculation of the baryon asymmetry.
If the spontaneously broken symmetry is
associated with the baryon number, the baryon asymmetry
generated by the decay of the PNGB field
was calculated [6] as $|\dot n_B| = \Gamma f^2 |\dot \theta|$
which gives
$$
|\Delta n_B| = \Gamma f^2 |\Delta \theta|
\eqno (3.6)
$$
Our first caveat is with regard to energy conservation.
The initial energy density of the field $\theta$ which creates the
baryons is $\rho_\theta(t_i) \sim f^2 m^2 \theta_i^2$.
At the end some of this energy density has been converted to
baryons, with energy density
$\rho_B(t_f) > n_B E_B $ where $n_B$ is the density of the
baryonic charge and $E_B \sim m$ is the characteristic
energy of the produced fermions.  Clearly it must be
true that $\rho_B(t_f) < \rho_\theta (t_i)$.  If we
were to use Eq. [3.6] we would see that this requires
$\Gamma < \Delta \theta m$. From the definition of $\Gamma$
we see that this is satisfied for small values of
coupling constant $g$ as long as $\Delta \theta$ is
not too small; for extremely small values of $\Delta \theta$,
this relationship can never be satisfied.

Our second caveat is as follows:  in making the identification
$|\dot n_B| = \Gamma f^2 |\dot \theta|$,
one is equating an operator equation (2.11)
with a vacuum averaged equation (3.4).  Indeed we started
with the operator equation which in our case looks like
$\ddot \theta + m^2 \theta = \dot n_B/f^2$.  Comparison
with Eq. (3.4) gives the identification mentioned above.
However, Eq. (3.4) is not an operator equation but obtained by the
vacuum averaging of the operator equation (2.11).  As we
have seen the average value $\langle \dot n_B \rangle$
is not just $\Gamma f^2 \dot \theta$ but a more
complicated expression (3.3).  This issue should be looked at
further.

Note that in the case of the spontaneous symmetry breaking without
any explicit one, when $U' (\theta ) =0$, the operator equation of
motion reads $f^2 \ddot \theta = \dot n_B $ and
$\Delta n_B = f^2 \Delta \dot \theta $ (as was mentioned previously).
In this case the characteristic energy of the produced fermions is
$\dot \theta$ and their energy density is of the second order
in $\dot \theta$. This agrees with the energy density of the creating
field $\theta$.  Thus the final fermion energy is indeed
consistent with the original energy in the $\theta$ field.

The third caveat is that Eq. (3.3) reduced to Eq. (3.4)
only in the approximation that the lower limit of
integration was taken to be $- \infty$.
This approximation is good as long as there
are many oscillations over the course of the integral.
For the case of inflation the field oscillates
many times during the reheating period, and thus
this approximation is probably reasonable.

\bigskip
\centerline{\bf IV. Further Complications:
i) Curved Spacetime (Expansion of the Universe),}
\centerline{\bf ii) Nonzero Fermion Masses, and iii) Parametric
Resonance}

{\it Curved Spacetime:}
So far all our results have been for flat spacetime.
In order to include the effects of the expansion of the
universe, we now generalize to curved spacetime:
$$
S = \int d^4x \sqrt{-g}
\bigg[ {1\over 2} f^2 (\partial_{\mu} \theta)
(\partial_{\nu} \theta) g^{\mu\nu}
+ i \bar Q \nabla_\mu \Gamma^\mu Q +
$$
$$i \bar L \nabla_\mu \Gamma^\mu L +
\partial_{\mu} \theta \bar Q \Gamma_\nu Q g^{\mu\nu} - m_Q \bar Q Q
-m_L \bar L L + gf(\bar Q L + \bar L Q)
- U(\theta)\bigg] \, .
\eqno(4.1)$$
We will consider Friedmann-Robertson-Walker metrics and
work in conformal time, $ds^2 = a^2 (d \tau^2 - d \vec{x}^2)$.
Since $g_{\mu\nu} = a^2 \eta_{\mu\nu}$, where $\eta_{\mu\nu}$
is the flat spacetime metric, by making this conformal
transformation in the Lagrangian we can reduce the metric
to the flat one.
With this transformation, we can use the ordinary
Minkowski space quantization for the fields and the usual
Green's functions.  To simplify we redefine
the fermion fields as $\psi \rightarrow \psi/a^{3/ 2}$.
Note that the gamma matrices in curved space-time $\Gamma^\mu$
are now
transformed to normal Dirac matrices $\gamma^\mu$.
Then the action is
$$
S = \int d^4x \bigg[ {1 \over 2} f^2 a^2 \partial_{\mu} \theta
\partial^{\mu} \theta + i \bar Q \partial_{\mu} \gamma^\mu Q +
i \bar L \partial_{\mu} \gamma^\mu L
+ \partial_{\mu} \theta \bar Q \gamma^\mu Q
$$ $$- m_Q a \bar Q Q
- m_L a \bar L L + gfa(\bar Q L + \bar L Q) - a^4 U(\theta)\bigg]
\, , \eqno(4.2)$$
where the summations are now done using $\eta_{\mu\nu}$.
With this action the equations of motion are
$$f^2 \partial_{\mu}(a^2 \partial^{\mu} \theta) +
\partial_{\mu}(\bar Q\gamma^\mu Q) + U'(\theta) a^4 = 0 \, , \eqno(4.3)$$
$$i \partial_{\mu} \gamma^{\mu} Q + \partial_{\mu} \theta
\gamma^{\mu} Q - m_Q a Q = -gfaL \, , \eqno(4.4)$$
and
$$i \partial_{\mu} \gamma^{\mu} L - m_L a L
= -gfaQ \, . \eqno(4.5)$$
For the case $m_Q = m_L =0$, we know the fermion Green's
functions (for nonzero masses there is the complication
that $m a(\tau)$ is time-dependent).  In the massless case
we can repeat the derivation done in flat spacetime previously
and find that the semiclassical equation of motion for the
Goldstone field is
$$\langle
\partial_{\mu}(a^2 \partial^{\mu} \theta) +
U'(\theta) a^4/f^2 \rangle =
$$ $$ - {4g^2 \over \pi^2} a(\tau) \int_0^\infty
d \omega \omega^2 \int_{-\infty}^0 d{\tau}' {\rm sin}(2 \omega {\tau}')
a(\tau + {\tau}') {\rm sin}[\theta(\tau + {\tau}') - \theta(\tau)]
\, . \eqno(4.6)$$
Of course, the lower limit of the ${\tau}'$ integral should
really be some initial time rather than $-\infty$.

{\it Nonzero Fermion Masses:}  Here we will consider the
modifications to the flat spacetime case when the fermion
masses are nonzero.  To the Lagrangian in Eq. (2.7) we
add terms $- m_Q \bar Q Q - m_L \bar L L$.  The equations
of motion (2.9) are modified to
$$
 i \gamma^\mu \partial_\mu Q + \partial_\mu \theta \gamma^\mu Q
-m_Q Q + gf L = 0 \,
\eqno(4.7a) $$ and
$$
i \partial_\mu(\bar Q \gamma^\mu) - \partial_\mu \theta  \bar Q
\gamma^\mu +m_Q \bar Q - g f \bar L = 0 \, .
\eqno(4.7b)$$
Equations (2.10) and (2.11) are unchanged.
Again, we wish to calculate Eq. (2.17), and, again,
$\langle \bar L(x) G_Q(x,y) L(y) \rangle $ is given by Eq. (A.2).
This time we keep the masses $m_Q$ and $m_L$ nonzero.
After we perform the $\int d^3y$ integral, there is
a term inside the remaining integrals:
$(l \cdot p-m_Q m_L) / (l^2 - m_Q^2) = (E_l E_p - \vec{l}^2 - m_Q m_L)
/( E_l^2 - \vec{l}^2 - m_Q^2).$  There are poles at
$E_l = \pm \sqrt{\vec{l}^2 + m_Q^2}$.  Again, only the $E_l < 0$
part gives a nonzero contribution.
Our result is
$$
{\cal D}^2 \theta + U'(\theta)/f^2 =
$$ $$- {g^2 \over \pi^3}
\int d^3l \int_{- \infty}^0 dt'
{\rm sin}[(E_p + E_l) t'] {\rm sin}[\theta(t+t') - \theta(t)]
{E_l E_p + \vec{l}^2 + m_Q m_L \over 2 E_p E_l}
. \eqno(4.8)$$
Here $E_l^2 = \vec{l}^2 + m_Q^2$ and $E_p^2 = \vec{l}^2 + m_L^2$.
Eq. (4.8) reduces to Eq. (2.21) when $m_Q = m_L = 0$.

For the case of $\dot \theta = const$ (Case I considered above),
we can see that only particles with masses
$ m_1 + m_2 < \dot \theta $
can be produced in perturbation theory, according to Eq. (4.8).
If we do the $t'$ integral in $\dot \theta = const$ case,
we get $\delta(\dot \theta - E_1 - E_2)$.  For $\dot \theta
< m_1 + m_2$ this can never be satisfied, there is no
particle production, and $\ddot \theta \equiv 0$ exactly.
For $\dot \theta > m_1 + m_2$, this delta function can
be satisfied for some momentum, and particles are produced.

The question remains what happens if one looks beyond
perturbation theory, particularly in an
expanding universe.  In the case of $e^+ e^-$ production
by a slowly varying electric field, a nonperturbative
contribution to particle production exists for the case where
the oscillation frequency $\omega$ is less then the electron mass $m_e$;
the result is that the production is exponentially suppressed
(the effect $\propto \exp[-const(m_e/\omega)]$) but nonzero.
Although we have not found such contributions here, they may
exist (see also ref. [3]).

{\it Parametric Resonance:}  Recently Kofman, Linde, and
Starobinski [7] (see also the work of Shtanov,
Traschen, and Brandenberger) [8]
noticed that parametric resonance may greatly enhance
the production of bosons during reheating in inflation;
one can interpret this as the formation of a Bose condensate.
In this paper we have been primarily considering the
production of fermions, for which there is no parametric
resonance. However, we should also consider the production
of $\theta$ bosons themselves by the classical $\theta$-field.

We will consider particle production during the reheating
phase of Natural Inflation.  As our potential for the
PNGB field we take $U(\theta) = \Lambda^4(1-{\rm cos}\theta)$.
Then the mass of the PNGB field is $m^2 = \Lambda^4/f^2$.
For Natural Inflation, $f \sim m_{pl}$ and $\Lambda \sim
10^{15}$ GeV, so that $m \sim 10^{12}$ GeV.
We will neglect coupling to fermions in our study of
the possibility of resonance, and include only coupling
of the field to itself.
At first, we will neglect expansion of the universe and
see that parametric resonance does indeed exist for
a few particular choices of wavenumber.  Then
we will include expansion and argue why we believe
that the resonance disappears.  We have not performed
a complete analysis of the equation in the case of
expansion, but for the case of Natural Inflation the
arguments are quite robust.
We suspect that the disappearance of the resonance
in an expanding universe may happen in other cases as well.
We leave investigation of this effect in other cases
to future work.

In flat spacetime, the equation of motion for
the PNGB is then $\ddot \theta + U'(\theta)/f^2 = 0$.
For our choice of potential, this becomes
$\ddot \theta + m^2 {\rm sin}\theta = 0$.
For small oscillations about $\theta = 0$,
we find the solution to this simple equation
to be
$$\theta_0(t) = \theta_{i} {\rm sin}mt \, . \eqno(4.9)$$
Here, subscript $i$ refers to initial value
of the unperturbed solution.  Following the approach
of references [7] and [8],
we will now add fluctuations
$\theta = \theta_0 + \delta \theta$.
We will keep terms to first order in $\delta \theta$.
Then $U' = \Lambda^4 {\rm sin}(\theta_0 + \delta \theta) =
\Lambda^4 [ {\rm sin} \theta_0
{\rm cos}(\delta \theta) + {\rm sin}(\delta \theta) {\rm cos}\theta_0]
\approx \Lambda^4 [{\rm sin}\theta_0 + \delta \theta {\rm cos}\theta_0]$
in the small angle approximation.
To first order in $\delta \theta$, after performing a
Fourier transform and substracting the zeroth order
piece, the equation of motion for $\delta \theta$ is then
$$\delta \ddot \theta +\Bigg[ {k^2} + {\rm cos}\theta_0 m^2\Bigg] \delta
\theta = 0 \, . \eqno(4.10)$$
Expanding around $\theta_0 = 0$ and using Eq. (4.9), we have
$$\delta \ddot \theta + [k^2 + m^2 - {m^2 \over 2} \theta_i^2 {\rm sin^2}mt]
\delta \theta = 0 \, . \eqno(4.11) $$
We define $y = 2 m t$ and
write ${\rm sin}^2 mt =  (1 - \cos 2mt)/2$.
Eq. (4.11) can then be written
$$
{d^2 \over dy^2} \delta \theta + \delta \theta \Bigg[ {k^2 \over 4 m^2} +
{1 \over 4} - {1 \over 16} \theta_i^2 + {1 \over 16} \theta_i^2
 \cos y \Bigg] =0
\, . \eqno(4.12)$$

The standard form of the Mathieu equation is
$$
{d^2 z\over dy^2} + (A + 2 \epsilon \cos y)z = 0 \, . \eqno(4.13)$$

Our flat spacetime Eq. (4.12) is of the form of the Mathieu equation
with $A = 1/4 - \theta_i^2/16 + k^2/4 m^2$
and $2 \epsilon = \theta_i^2/16$.  Since we made the
small angle approximation earlier on, we've assumed
$\theta_i < 1$, i.e. $\epsilon \ll A < 1$.
The Mathieu equation has been studied in great depth.
It is known that for $A=1/4 +A_1 \epsilon$ and $\epsilon \ll 1$,
there is no
instability if $|A_1 |>1$ (In our case $|A_1| =2$ for
$k=0$.)
Therefore, for $k = 0$, there is no instability.
However, for particular values of nonzero $k$
(values for which $A \sim n^2/4$ where $n$ is an integer)
there are indeed regions of resonance
with $ \delta \theta $ growing exponentially in time.
We refer the reader to literature on the Mathieu equation
to see these regions.

However, now let us include the expansion of the universe.
To simplify we will neglect here the interaction with fermions.
Without fermions it is convenient to work in terms of the
physical time $t$ with the interval $ds^2 = dt^2 -
a^2(t) d{\vec r}^2$.
The relevant change from the nonexpanding case will not
only be the additional term $3 H \dot \theta$
in the equation of  motion; rather the important features are
i) the redshifting of the lengthscales of the perturbations
($k \rightarrow k/a$)
and ii) the fact that the unperturbed solution
$\theta_0$ is now different.
The equation of motion (again, neglecting the fermion effects)
becomes  $\ddot \theta + 3 H \dot \theta + U'(\theta)/f^2 = 0$.
For our choice of potential, this equation becomes
$\ddot \theta + 3 H \dot \theta + m^2 {\rm sin}\theta = 0$.
Again, we take the small angle
approximation ${\rm sin} \theta \sim \theta$.
The unperturbed solution for the matter dominated expansion is
$$
\theta_0(t) = {\theta_i {\rm sin}(mt +\delta)\over mt} \, .
\eqno(4.14)$$
The factor of $1/t$ in the denominator will prove
to be an important feature of the expansion.
We assumed here that the $\theta$-field started to oscillate when
the Hubble parameter, $H=2/3t$, was close to the value of the
mass of the field $m$, as is usually the case for inflation.
It fixes the
initial value of time, $t_i \sim 1/m$.

This time we take $y = mt$.
The equation for the fluctuations becomes
$${d^2\over dy^2} \delta\theta + {3 H\over m} {d \over dy}
\delta \theta + \Bigg[{k^2 \over m^2 a^2} + 1 - {1 \over 2} \theta_i^2
{{\rm sin}^2 y \over y^2}\Bigg] \delta \theta = 0 \, .
\eqno(4.15)$$
To eliminate the $\delta \dot \theta$ term we define
$\theta(t) = \bar \theta(t) / a^{3/2}$.
During the reheating portion of inflation, the universe
is matter dominated and we take $a \propto t^{2/3}$ (of course
after reheating the universe is radiation dominated).
With this matter dominated expansion, Eq. (4.15) becomes
$$
{d^2\over dy^2} \delta \bar \theta +
\Bigg[{k^2 \over m^2 a^2} + 1 -  {\theta_i^2 \over 2}
{{\rm sin}^2 y \over y^2}\Bigg] \delta \bar \theta = 0 \, .
\eqno(4.16)
$$
As before, we use ${\rm sin}^2 y = [1 - {\rm cos}(2y)]/ 2$
to write the equation in the form closest to the Mathieu
equation.  This time it is not exactly the Mathieu equation
because of the time dependence in the denominator of $k^2/a^2$
and because of the factor of $1/y^2$ that came from Eq. (4.14).

We have numerically integrated Eq. (4.12) without
universe expansion
and Eq. (4.15) with the expansion
to see what happens to the resonance. As we expected the solutions of
eq. (4.12) show the resonance behavior for a particular region
of parameters while solutions of eq. (4.16) do not resonate.
The resonance might be excited if the oscillations of $\theta$
began when $H\sim 1/t_i \ll m$ (not what usually happens
in inflation).  We did not perform a numerical study of the
entire range of parameter space, and in principle could
have missed the particular choices of $k/a$ that do resonate.
Hence we proceed here with a simple analytic discussion.

One can see from simple analytic arguments that Eq. (4.16)
is unlikely to lead to resonance. When one includes expansion,
there are two effects that reduce the instability.
First, the redshift of the wavenumber, $k \rightarrow k/a$,
quickly moves any wavenumber that happens to be in a resonance
band, out of the resonance region.  In other words,
if at one time there is an instability on some lengthscale,
shortly afterwards this lengthscale has redshifted
to a value for which there is no instability.
Thus it is difficult to see
how there could be exponential increase in particle production
(numerically we could get factors of a few, not of $10^5$).
Second, in the long time limit, $y \gg 1$, both the first
and third terms inside the brackets become small.  Then the
equation is simply a harmonic oscillator equation with
oscillating rather than unstable solutions.  The third term,
whose negative sign could make it responsible for resonance,
becomes unimportant for times $t > \theta_i/2m$. Since the
frequency of oscillations $m$ is usually faster than the frequencies
corresponding to other
relevant timescales, such as the timescale of reheating in
inflation, the third term quickly becomes unimportant.
There may be enhancement in the first oscillation or two,
but it is probably not very large (as above, this assumes
that the oscillations began when $H \sim m$, i.e. $y \sim 1$).
Instead the solutions quickly become oscillatory,
with amplitudes at most slightly larger than those in
the non-expanding case.
Again, we have not performed a complete analysis of the
equation (4.16), but we have argued why we
believe the resonance effects are not very strong here.

\vskip .125 in
\centerline{\bf V. Baryogenesis in Natural Inflation}

The inflationary universe model [1] provides an elegant means of
solving several cosmological problems, including the horizon problem,
the flatness problem, and the monopole problem.  In addition, quantum
fluctuations produced during the inflationary epoch may provide the
initial conditions required for the formation of structure in the
universe.  During the inflationary epoch, the energy density of the
universe is dominated by a (nearly constant) vacuum energy term
$\rho \simeq \rho_{vac}$, and the scale factor $R$ of the universe
expands superluminally (i.e., ${\ddot R} > 0$).  If the time interval
of accelerated expansion satisfies $\Delta t \ge 60 R/ {\dot R}$, a
small causally connected region of the universe grows sufficiently
to explain the observed homogeneity and isotropy of the universe,
to dilute any overdensity of magnetic monopoles, and to flatten the
spatial hypersurfaces (i.e., $\Omega \rightarrow 1$).

The model of Natural Inflation [2] was proposed to provide
a natural explanation of the required flatness of the potential
in inflation.
The flatness is achieved by mimicking the axion physics
described earlier.  The inflaton is a PNGB field and two
different mass scales describe the height and width
of the potential.  The model has several nice features,
including the possibility of extra large scale power
in the density fluctuations, negligible production of
gravitational waves, and possible tie-ins to particle
physics models under consideration.

Here we want to consider an idea for baryogenesis
during Natural Inflation.  More standard ideas such
as a) reheating to above the baryogenesis temperature
(e.g. electroweak) or b) baryon violating decays
have already been considered [2].
Instead, here we consider a model of baryogenesis in
which the baryon number is produced as the inflaton
is rolling down its potential. In this paper
we will merely suggest the idea, and leave study
of the implementation of the idea for future work.

For this particular
idea to work, the inflaton would have to carry baryon
number.  If the inflaton rolls
clockwise down the hump in the Mexican hat, then
baryons are produced; if the inflaton rolls counterclockwise
down the hump, then antibaryons are produced.  In different
regions of the universe, there will be these two different
kinds of behavior.  Any one region will be blown up to become
very large by the inflation.  Thus our observable universe, which lies
inside one of these regions, had a fifty/fifty chance
of being made primarily of baryons/antibaryons.  CP
violation is not explicitly required in the Lagrangian,
as the sign of the baryon number
is determined by the initial conditions, namely
the direction of the roll of the field.  These
ideas are very similar to those of [5,6].

The reheating temperature in this scenario could
be very low. In particular, if $T_{reheat} < T_{electroweak}$,
sphalerons do not erase any baryon asymmetry
generated during inflation.  Of course we also need
to return to the standard evolution of the universe
at a high enough temperature for nucleosynthesis;
i.e. $T_{reheat} > T_{nucleosynthesis}$. It may be
a nice feature to have a very low reheating temperature,
as many inflationary models are very constrained by
the requirement of a high reheat temperature.

In order for this to work, the $\Phi$ field must carry
baryon number.  The current that is explicitly broken
by instantons or by whatever else provides the tilt
(the cosine potential) cannot be orthogonal to baryon number.
In that case the baryon number of the $\Phi$ field will be
proportional to the angular momentum as the inflaton
rolls down: one direction of roll will correspond
to baryon production and the other to antibaryon production.
The baryon current carried by the $\Phi$ field is
$J^\mu = i[\Phi \partial^\mu \Phi^* -
\Phi^* {\partial}^\mu \Phi]$.  Since $\langle \Phi \rangle
= f e^{i\theta}$, the baryon number density
$\langle n_B\rangle \equiv \langle J_0 \rangle
= f^2 \dot \theta$,
namely the angular momentum of the
two-dimensional mechanical motion of the $\Phi$ field in
in the plane $(Re\Phi, Im\Phi)$.  As an
example, in the Lagrangian in Eq. (2.7), we have
considered a symmetry whereby $\Phi$ and $Q$ transform
whereas $L$ does not.  Thus $\Phi$ and $Q$ could carry
baryon number while $L$ does not.  $Q$ and $L$ would
not be ordinary quarks and leptons; rather they would
be hidden sector particles that could be made to couple
to quarks and leptons in such a way that $Q$ carries
baryon number while $L$ does not.

There are many constraints that such a model must satisfy.
One must be careful about the quantum numbers carried
by the various fields: namely SU(3) color, the gauge
group that became strong at scale $\Lambda$ and produced
the cosine potential in the first place, and baryon number.
One must also ensure that the present day violation of
baryon number predicted for ordinary matter is not
in excess of observations.
Also, we do not want the only decay mode to be to
baryonic matter of our universe.  Somehow there must
be nonbaryonic decay modes
or decays to baryons that remain
in the hidden sector.
Simultaneously satisfying all these constraints is
difficult. However,
we have by no means exhausted all the possiblities, and
leave this investigation to future work should the idea
prove promising enough.
As indicated near the end of Section III above,
the calculation of the baryon
number produced in such a model will be
performed in future work.

\vskip .125in
\centerline{\bf VI. Conclusions}

As a Nambu-Goldstone boson $\theta$ moves in a potential, it
can produce fermions that it couples to.  A semiclassical
calculation of particle production
was performed.  The backreaction of quantum fermions
on the evolution of a classical $\theta$ was calculated
for a specific simple model, to provide a general framework
within which one can calculate production of other particles
as well. The primary results of our calculations in flat
spacetime are Eqs. (2.21) and (2.22).  Generalization to
curved spacetime with massive fermions was discussed.
We argued that enhanced production of bosons due to parametric
resonance is probably not important here in an
expanding universe; a more general investigation of
the effects of expansion on resonance is warranted in the future.
We are especially interested
in the model of Natural Inflation, in which the inflaton is
a pseudo Nambu Goldstone boson.  It
may be possible for the inflaton to
create baryon asymmetry at the exit
from inflation simultaneously with the universe reheating.

\newpage
\vskip 1.0truein
\centerline{\bf APPENDIX A:}
\centerline{\bf CALCULATION OF SEMICLASSICAL EQUATION FOR $\theta$ FIELD}
\nobreak \bigskip

We wish to calculate the right hand side of Eq. (2.17).
Using the quantization in Eq. (2.19), we find that
$$\langle \bar L(x) G_Q(x,y) L(y) \rangle = $$ $$\sum_s \sum_{s'}
\int {d^3p d^3p' \over (2 \pi)^3} {m_L \over \sqrt{E_p E_{p'}}}
\bar v_{p'}^{s'} G_Q(x,y) v_p^s e^{-ip' \cdot x}
e^{ip \cdot y} \langle d^{s'}_{p'}
d_p^{s{\dagger}} \rangle$$
$$= \sum_s \int {d^3p \over (2 \pi)^3} {m_L \over E_p}
\bar v^s_p G_Q(x,y) v^s_p e^{-ip \cdot (x-y)} \, , \eqno(A.1)$$
where the second equality follows since
$\langle d^{s'}_{p'} d_p^{s{\dagger}} \rangle = \delta_{s s'}
\delta^3(\vec{p} - \vec{p}')$.
Using $$G_Q(x,y) = i \int {d^4l \over (2 \pi)^4} {exp[-il \cdot (x-y)]
\over l_\mu \gamma^\mu - m_Q + i \epsilon} \eqno(A.1a)$$
and $\sum_s v^s_p \bar v^s_p = (p_\mu \gamma^\mu -m_L) / 2 m_L$,
we find
$$\langle \bar L(x) G_Q(x,y) L(y) \rangle =$$
$$\int {d^3p \over (2 \pi)^7} \int {d^4l \over 2 E_p}
\Bigg[{i \over l^2 + m_Q^2}\Bigg]
e^{[i(p+l) \cdot (y-x)]} {\rm Tr}[(l_\mu \gamma^\mu
+m_Q)(p_\mu \gamma^\mu -m_L)] \, .
\eqno(A.2)$$
In the massless limit considered in Section II, the trace becomes
$Tr[l_\mu p_\nu \gamma^\mu \gamma^\nu] = 4l \cdot p$ and we have Eq. (2.20).
The calculation of $\langle \bar Q(x) G_L(x,y) Q(y) \rangle$
is similar.

Using Eq. (2.20), we now have terms on the right hand side of
Eq. (2.17) such as
$$I = -i g^2 \int d^4y e^{[i \theta(x) - i \theta(y)]} \int {d^3p
\over (2 \pi)^7} {d^4l \over E_p l^2}
e^{[i(p-l) \cdot (y-x)]} l \cdot p + h.c.
\, , \eqno(A.3)$$
where we have taken $l \rightarrow -l$ compared to previous
expressions.  We now write $l^\mu =(E_l, \vec{l})$ and
$p^\mu = (E_p, \vec{p})$
so that $l \cdot p = E_l E_p - \vec{l} \cdot  \vec{p}$.
Henceforth we will assume no spatial gradients in the
$\theta$ field, i.e. $\theta = \theta(t)$ only.  We will
now use $\int d^3y e^{[i(\bar p - \vec{l}) \cdot \vec{y}]} =
(2 \pi)^3 \delta^3(\vec{p} - \vec{l})$
to write
$$I = -ig^2 \int dt_y e^{[i \theta(t_x) - i \theta(t_y)]}
\int {d^4l \over (2 \pi)^4} {d^3p \over E_p}
e^{[i(E_p - E_l)(t_y - t_x)]} \delta ({\vec p} - {\vec l})
{E_l E_p - \vec{l}^2 \over
E_l^2 - \vec{l}^2} + h.c. \, \eqno(A.4)$$
We now do $\int dE_l$ and find a nonzero contribution
from the pole at $E_l = - |\vec{l}|$.
Note that we take the retarded Green's function,
which gives nonzero result only for $t_y < t_x$.
We have
$$I = {g^2 \over 2 \pi^2} \int_0^\infty E_l^2 dE_l
\int_{-\infty}^{t_x} dt_y e^{[2i E_l(t_y - t_x)]}
e^{[i \theta(t_x) -i \theta(t_y)]} + h.c. \,  \eqno(A.5)$$
Adding all terms of this form that contribute to the right
hand side of Eq. (2.17), relabeling $E_l$ as $\omega$,
and defining $t' = t_y - t_x$,
we find
$$
\langle {\cal D}^2 \theta + U'(\theta)/f^2 \rangle
= - {4g^2 \over \pi^2}
\int_0^\infty d \omega \omega^2 \int_{- \infty}^0 dt'
{\rm sin}(2 \omega t') \, {\rm sin}[\theta(t+t') - \theta(t)] .
\eqno(A.6)$$
This is the result quoted in Eq. (2.21).

We can rewrite this in the form given in Eq. (2.22) if
we now perform the $\omega$ integral:
$$\int_0^{\infty} d\omega \omega^2 {\rm sin}(2 \omega t')
= - {1 \over 4} {\partial^2 \over \partial t'^2}
\Bigg[\int_0^\infty d\omega {\rm sin}2 \omega t' \Bigg]
$$ $$= \lim_{w \rightarrow \infty} {\partial^2 \over \partial t'^2}
\Bigg[{1 \over 2 t'} \Bigg({\rm cos}2 w t' - 1 \Bigg)
\Bigg]\, . \eqno(A.7)$$
Next we do the $t'$ integral by parts.
We will use the notation $\Delta \theta = [\theta(t+t') - \theta(t)] $.
The nonvanishing contribution is
$${1 \over 8} \lim_{w \rightarrow \infty} \int_{-\infty}^0 dt'
{\partial^2 \over \partial t'^2}
\Bigg[{1 \over  t'} \Bigg({\rm cos}2 w t' - 1 \Bigg)
\Bigg] {\rm sin}\Delta \theta $$
$$= -{1 \over 8} \lim_{w \rightarrow \infty}  \int_{-\infty}^0 dt'
{\partial \over \partial t'}
\Bigg[{1 \over  t'} \Bigg({\rm cos}2 w t' - 1 \Bigg)
\Bigg] {\partial \over \partial t'} {\rm sin}\Delta \theta $$
$$= {1 \over 8} \lim_{w \rightarrow \infty} \int_{-\infty}^0 dt'
\Bigg[{1 \over  t'}
\Bigg({\rm cos}2 w t' - 1 \Bigg)
\Bigg] {\partial^2 \over \partial t'^2} {\rm sin}\Delta \theta\,
, \eqno(A.8)$$
where surface terms have all vanished.  We now perform
the derivative on ${\rm sin} \Delta \theta$.
Then Eq. (A.6) becomes
$$\langle {\cal D}^2 \theta + U'(\theta)/f^2 \rangle =
- {g^2 \over 2\pi^2} \lim_{w \rightarrow \infty}
\int_{-\infty}^0 dt' \Bigg[{{\rm cos}2 w t' - 1 \over
 t'}\Bigg] \Bigg[\ddot \theta(t+t') {\rm cos}\Delta \theta
- \dot \theta^2(t+t') {\rm sin}\Delta \theta\Bigg]\, , \eqno(A.9)$$
This is the result quoted in Eq. (2.22)

\vskip 1.0truein
\centerline{\bf APPENDIX B:}
\centerline{\bf CALCULATION OF MASS RENORMALIZATION TERM}
\centerline{\bf FOR THE CASE OF SMALL OSCILLATIONS AROUND THE MINIMUM}
\nobreak \bigskip

To obtain Eq. (3.4), we must calculate the following term:
$$m_R^2 \theta_0  {\rm cos}m_R t \int_{- \infty}^0
{dt{'} \over t{'}} [1 - {\rm cos}2 w t'] {\rm cos}m_R t' = $$
$$m_R^2 \theta_0  {\rm cos}m_R t
\int_{- \infty}^0 {dt' \over 2t'} \Bigg[{\rm cos}m_R t' -
{\rm cos}\alpha t' + {\rm cos}m_R t' - {\rm cos}\beta t' \Bigg]$$
$$= m_R^2 \theta_0  {\rm cos}m_R t
\int_{- \infty}^0 {d t' \over t'} \Bigg[{\rm sin}(m_R+w) t'
{\rm sin}wt' + {\rm sin}w t'
{\rm sin}(w-m_R)t'\Bigg] \, , \eqno(B.1)$$
where $\alpha = 2w + m_R$ and $\beta = 2 w -m_R$.
After doing the integral, we find that this term is
$$-{m_R^2 \theta_0 \over 2} {\rm cos}m_R t
\Bigg[{\rm log}\Bigg({2w+m_R \over m_R}\Bigg) + {\rm log}
\Bigg({2w-m_R \over m_R}\Bigg)\Bigg] \,
. \eqno(B.2)$$
In the limit $w \rightarrow \infty$,
this becomes
$$- m_R^2 \theta_0  {\rm cos}m_R t
\, \, {\rm log}(2 w/m_R) \, , \eqno(B.3)$$
a logarithmically divergent term that renormalizes the mass.

\bigskip
\bigskip
\bigskip
\centerline{\bf Acknowledgements}
\medskip
\nobreak

K.F. would like to thank Liz Jenkins, Salman Habib, Glennys Farrar,
and especially Steve Selipsky and Fred Adams for useful conversations.
K. F. was supported by the NSF Presidential Young Investigator Program.
A. Dolgov is grateful for the hospitality
of the Department of Physics of the University
of Michigan where this work was started.

\newpage
\vskip 1.0truein
\centerline{\bf REFERENCES}
\nobreak \vskip 0.10truein

\item{[1]}{A. H. Guth, {\it Phys. Rev.} D {\bf 23}, 347 (1981).}

\item{[2]}{K. Freese, J. A. Frieman, and A. V. Olinto,
{\it Phys. Rev. Lett.} {\bf 65}, 3233 (1990);
F. C. Adams, J. R. Bond, K. Freese, J. A. Frieman,
and A. V. Olinto, {\it Phys. Rev.} D {\bf 47}, 426 (1993).}

\item{[3]}{A. Dolgov and Kirilova, {\it Yad. Fiz.} {\bf 51},
273 (1990); {\it Sov. J. Nucl. Phys.} {\bf 51}, 172 (1990).}

\item{[4]}{L.F. Abbott, E. Farhi, and M.B. Wise,
{\it Physics Letters} {\bf 117B}, 29 (1982);
A. Albrecht, P.J. Steinhardt, M.S. Turner, and F. Wilczek,
{\it Phys. Rev. Lett} {\bf 48}, 1437 (1982);
A. D. Dolgov and A. D. Linde, {\it Phys. Lett.} {\bf 116B},
329 (1982); J. Traschen and R. Brandenberger, {\it Phys.
Rev. D} {\bf D42}, 2491 (1990); J. P. Paz, {\it Phys.
Rev. D} {\bf 42}, 530 (1990).}

\item{[5]}{I. Affleck and M. Dine, {\it Nuclear Physics B}
{\bf B249}, 361 (1985).}

\item{[6]}{A.G. Cohen and D.B. Kaplan, {\it Physics Letters}
{\bf B199}, 251 (1987).}

\item{[7]}{L. Kofman, A. Linde, and A. Starobinski, `Reheating
after Inflation',
University of Hawaii preprint UH-IfA-94/35.}

\item{[8]}{Y. Shtanov, J. Traschen, and R. Brandenberger,
`Universe Reheating after Inflation', Brown University
preprint BROWN-HET-957.}

\item{[9]}{H. Quinn and R. Peccei {\it Phys. Rev. Lett.}
{\bf 38}, 1440 (1977).}

\item{[10]}{As written, the Yukawa coupling $g \Phi \bar Q L + h.c.$,
would generate masses for the fermions $m_\psi \sim gf$.
These masses would be extremely large, in fact so large that
the Nambu-Goldstone boson would never be able to produce them.
However, this Yukawa coupling is just a schematic for the real
coupling.  As in the case of the axion models, the real Yukawa
coupling that gives a fermion its masses is a coupling
of the fermion to a different scalar, not directly the
PQ scalar.  For example, for the case of the Dine,
Fischler, Srednicki axion [11], the Yukawa couplings are with two
scalar doublets $\phi_u$ and $\phi_d$, which transform under
the chiral symmetry in the same way as the PQ field $\Phi$.
The VEVs of these other scalars are much lower in energy
scale than the VEV of the $\Phi$ field.  Thus the fermions
retain smaller masses.  The axion is then a linear combination
of all these scalars, but predominantly made
of the $\Phi$ component.  To mimic this physics without
including all the complications, we keep the simple Yukawa
coupling of Eq. (2.1), but take $g \ll 1$ to ensure proper
fermion masses.}

\item{[11]}{M. Dine, W. Fischler, and M. Srednicki,
{\it Physics Letters} {\bf B104}, 199 (1981).}

\item{[12]}{C. T. Hill and G. G. Ross, {\it Phys. Lett.}
{\bf B203}, 125 (1988); C. T. Hill and G. G. Ross, {\it Nucl. Phys.}
{\bf B311}, 253 (1979).}

\item{[13]}{S. Weinberg {\it Phys. Rev. Lett} {\bf 40},
223 (1978); F. Wilczek {\it ibid}, {\bf 40}, 279 (1978).

\bye

\newpage
\vskip1.0truein
\centerline{\bf FIGURE CAPTIONS}
\bigskip

\medskip
\noindent

\bye